\documentclass{article}
\usepackage{graphicx} 
\usepackage[super,sort&compress]{natbib}
\usepackage{booktabs}
\usepackage{array}
\usepackage[most]{tcolorbox}
\usepackage{float}
\usepackage{placeins}
\usepackage[margin=1in]{geometry}

\title{Machine learning cannot decarbonise the subsurface until it can be audited}

\author{Hannah P. Menke$^{1,2}$, Ahmed H. Elsheikh$^{1,2}$, Lingli Wei$^{3,4}$, \\
Nanzhe Wang$^{1,2}$, Andreas Busch$^{4,2}$}

\date{%
\small
$^{1}$Institute of GeoEnergy Engineering, Heriot-Watt University, Edinburgh, UK\\
$^{2}$Subsurface Energy Transition and Innovation Centre, Heriot-Watt University, Edinburgh, UK\\
$^{3}$SubsurfaceLabs, Abu Dhabi, UAE\\
$^{4}$Lyell Centre, Heriot-Watt University, Edinburgh, UK\\[0.5em]
Corresponding author: Hannah P. Menke, h.menke@hw.ac.uk\\[0.5em]
September 2026
}

\begin{document}

\maketitle

\begin{abstract}
Geological CO$_2$ storage, geothermal energy and underground hydrogen storage depend on irreversible or high-cost decisions made with sparse data and incomplete physical understanding. Machine learning (ML) increasingly informs these decisions, yet methods are usually judged on predictive accuracy rather than on whether their outputs can be trusted, traced and acted upon. We argue that subsurface ML should be designed as decision infrastructure with shared systems that carry calibrated uncertainty and provenance from observation to operating decision. We propose a decision-readiness checklist, an audit-trail workflow for monitoring updates, and benchmark designs that score models by the decisions they support. We identify validation needs shared across the three technologies, and the data-access and governance steps needed for ML to support regulatory and investment decisions.
\end{abstract}

\section{Introduction: subsurface decisions under uncertainty}
\subsection{What makes the subsurface different}
Geological CO$_2$ storage, geothermal energy and underground H$_2$ storage require irreversible or high-cost decisions about subsurface systems whose behaviour is only partly known. Sparse labelled data, non-unique inverse problems, and incomplete knowledge of physics and geology mean that several models may be consistent with the available evidence \cite{bergen2019machine}. Decisions must account for this range of plausible models \cite{ringrose2016reservoir}. Differences between basins, lithologies and operating conditions challenge the application of models beyond their training data. These characteristics, together with the environmental, regulatory and financial consequences of failure, make interpretability, uncertainty calibration and auditability essential \cite{pawar2015recent}. Calibration asks whether stated probabilities match observed frequencies.

Project evaluation weighs injectivity, heat recovery, and deliverability against containment \cite{busch2018migration}, pressure footprint \cite{de2026co2logix}, induced seismicity risk \cite{paluszny2024induced}, and energy requirements \cite{sgouridis2019comparative}. In injection-driven systems, hydro-mechanical coupling \cite{meng2026thermo,liu2025multi} can be the controlling constraint, requiring machine learning (ML) to respect geomechanical limits. Uncertainty in these quantities affects permitting risk, monitoring burden, and cost of capital.

These constraints can determine whether injection continues. At the In Salah carbon capture and storage (CCS) project in Algeria, analysis of reservoir, seismic and geomechanical data led to suspension of injection in June 2011 \cite{ringrose2013insalah}. At Pohang in South Korea, fluid injection for an enhanced geothermal system (EGS) project triggered a magnitude 5.5 earthquake in 2017 and the project was suspended \cite{ellsworth2019pohang}.

\subsection{Scope and contribution}
Here, subsurface decarbonisation technologies comprise geological CO$_2$ storage, geothermal energy including EGS, and underground H$_2$ storage in porous media. These technologies serve to decarbonise the wider energy system. We organise subsurface decarbonisation projects around a funnel of screening, appraisal, operations, closure, and stewardship, with measurement, monitoring and verification (MMV) spanning these stages. The funnel produces heterogeneous evidence (seismic, well and laboratory data, models, monitoring time series). At each handover, the uncertainty and provenance of this evidence should remain traceable into the next decision.

Earlier reviews have surveyed ML applications in subsurface systems by workflow stage \cite{yan2021ccus} and by method class \cite{wang2025comprehensive}. We ask what evidence ML must supply before an operator, regulator or investor can act on its outputs.

\subsection{From prediction tools to decision infrastructure}
ML outputs in subsurface decarbonisation can feed permit applications, monitoring plans, storage-liability transfers, and investment decisions that bind operators, regulators, insurers, and the public over decades. The shared, persistent, standardised and maintained systems that supply these outputs constitute \emph{decision infrastructure}. Many actors rely on these systems, including those who did not build them and cannot inspect their internals. Where ML occupies this role, calibration, provenance, audit and defined failure behaviour must shape method development from the outset.

The ML stack connects observations to decisions through four stages. \emph{Sensing} covers acquisition and quality control (QC). \emph{Inference} uses these measurements to interpret the subsurface and estimate parameters and states. \emph{Simulation} produces forecasts through physics-based models or surrogates. The \emph{decision} stage uses those forecasts to guide operations and monitoring under uncertainty. Across these stages, the objective is to combine heterogeneous data into traceable priors, calibrate uncertainty, enforce physics and geological rules, and optimise actions under explicit risk constraints \cite{ringrose2016reservoir,pawar2015recent}.

\begin{figure}[!htbp]
    \centering
    \includegraphics[width=\linewidth]{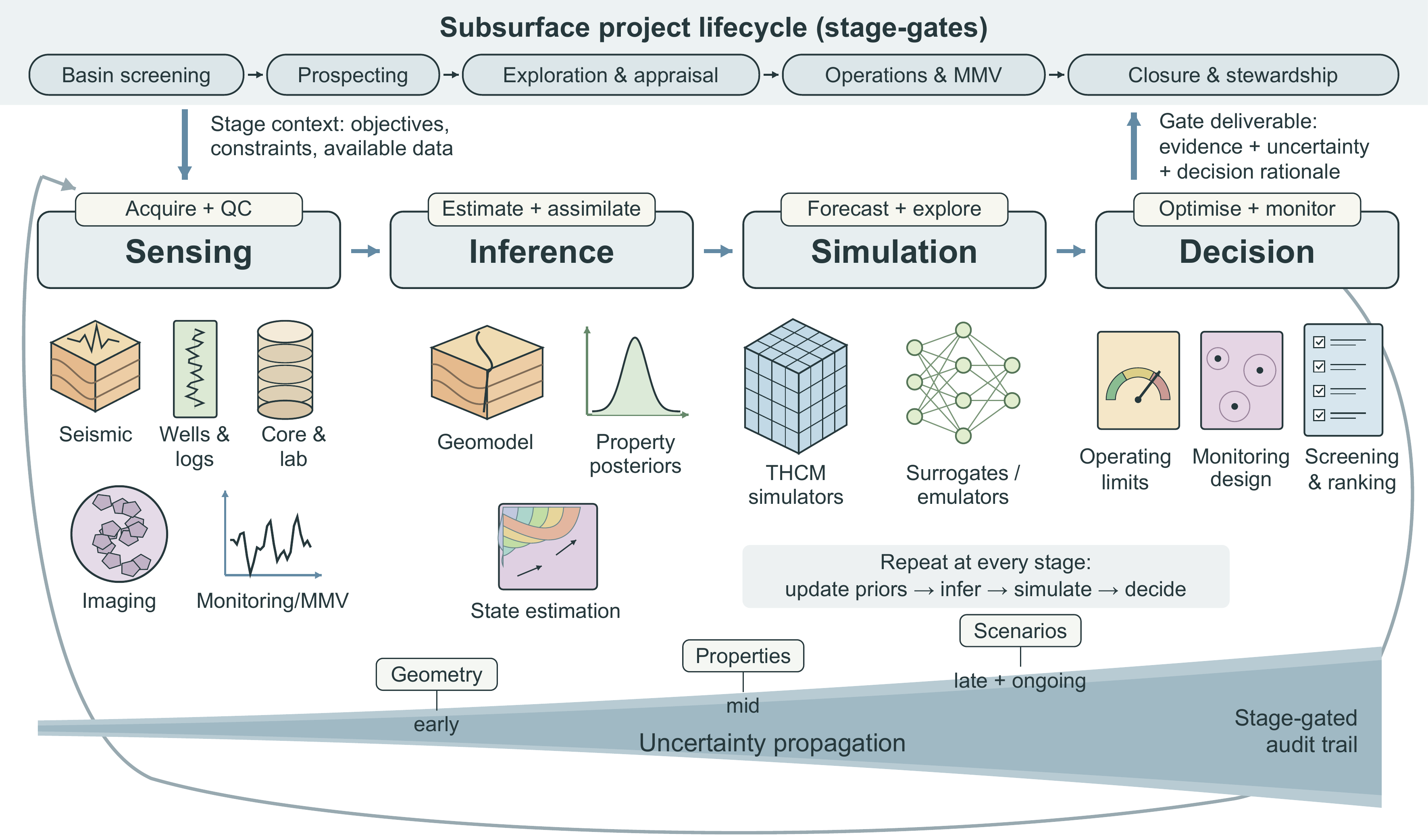}
    \caption{Funnel $\times$ Stack framing: the subsurface project funnel (basin screening $\rightarrow$ prospecting $\rightarrow$ exploration and appraisal $\rightarrow$ operations and MMV $\rightarrow$ closure and stewardship) repeated through the ML stack (sensing $\rightarrow$ inference $\rightarrow$ simulation $\rightarrow$ decision). The uncertainty ribbon highlights that uncertainty is generated at early funnel stages, shaped by new data and models as the project matures, and must ultimately be expressed as decision risk (operating envelopes and trigger criteria). THCM, thermo-hydro-chemical-mechanical.}
    \label{fig:ai_stack}
\end{figure}

The resulting decision outputs are operating envelopes and trade-off curves. Operating envelopes describe the set of operating conditions expected to satisfy specified physical and risk limits. Trade-off curves show how much one objective must be given up to improve another. When deciding which measurements to acquire, the \emph{value-of-information} criterion ranks candidates by their expected improvement in decision value, net of acquisition cost. Figure \ref{fig:ai_stack} shows how uncertainty evolves into decision risk through the project funnel.

\noindent\textbf{Funnel handovers that ML must make explicit.}
\begin{itemize}
\item Screening $\rightarrow$ Appraisal: convert regional analogues and seismic interpretations into \emph{structured priors} with uncertainty.
\item Appraisal $\rightarrow$ Operations: translate wells, logs, and special core analysis (SCAL) into \emph{decision-relevant uncertainty} (e.g.\ injectivity, pressure footprint, trapping, geomechanics).
\item Operations $\rightarrow$ Closure: convert MMV time series into \emph{audit-ready evidence} for containment, conformance, and long-term risk bounds.
\end{itemize}

\subsection{One decision problem, three different systems}
CO$_2$ storage, geothermal energy, and H$_2$ storage share the challenge of constrained decisions under uncertainty \cite{tester2021the,heinemann2021hydrogen}, but their physics and operating regimes differ, and a workflow tuned to one can fail on another.

Table \ref{tab:threesystems} compares operating regimes and coupled physics \cite{heinemann2021hydrogen,boreham2021hydrogen,meng2026thermo}. Section \ref{sec:mmv} examines the limits of validating long-term CO$_2$ containment on research timescales.

\begin{table}[!htbp]
\centering
\small
\caption{Shared structure and system-specific differences across the three subsurface decarbonisation technologies. Rows above the midline are largely shared; rows below differ materially and require system-specific treatment.}
\label{tab:threesystems}
\begin{tabular}{@{} >{\raggedright\arraybackslash}p{0.20\linewidth} >{\raggedright\arraybackslash}p{0.24\linewidth} >{\raggedright\arraybackslash}p{0.24\linewidth} >{\raggedright\arraybackslash}p{0.24\linewidth} @{}}
\toprule
 & \textbf{CO$_2$ storage} & \textbf{Geothermal / EGS} & \textbf{H$_2$ storage} \\
\midrule
Shared data types & \multicolumn{3}{p{0.72\linewidth}}{Seismic surveys, well logs, core and SCAL data, downhole pressure and temperature, microseismic monitoring, legacy reports and analogues.} \\
Shared ML methods & \multicolumn{3}{p{0.72\linewidth}}{Uncertainty-aware surrogates, ensemble data assimilation, structure-aware geological priors, hybrid physics-ML, multi-fidelity learning, value-of-information monitoring design.} \\
\midrule
Primary decision variables & Containment probability, pressure footprint, plume conformance & Thermal breakthrough time, induced seismicity risk, sustainable flow rate & Cyclic deliverability, recovery efficiency, containment under cycling \\
Dominant coupled physics & Multiphase flow, dissolution and residual trapping, pressure-driven geomechanics & Heat transport, fracture-dominated thermo-hydro-mechanical coupling & Hysteretic multiphase flow under cycling, microbial and geochemical H$_2$ reactions, cushion-gas mixing \\
Operational regime & One-way injection over decades; stewardship over centuries & Continuous circulation over years to decades & Seasonal to weekly injection-withdrawal cycles \\
System-specific data & Time-lapse (4D) seismic, surface deformation from interferometric synthetic aperture radar (InSAR), geochemical assurance monitoring & Tracer tests, temperature and flow logs, dense microseismic catalogues & Per-cycle pressure and gas-composition histories, well deliverability tests \\
System-specific ML needs & Long-horizon extrapolation without ground truth; plume tracking from 4D seismic & Fracture-network inference; seismicity forecasting for traffic-light systems & Hysteresis-aware surrogates; reactive-transport closures; rapid per-cycle model updating \\
Regulatory context & Storage directives and Class VI-type permits; long-term liability transfer \cite{eudirective2009,epaclassvi2010} & Seismicity thresholds and traffic-light protocols \cite{ellsworth2019pohang} & Gas-storage and safety regulation; standards still emerging \cite{heinemann2021hydrogen} \\
Validation opportunities & Operational pressure and plume forecasts; long-term containment extends beyond research timescales & Temperature and flow forecasts assessed against continuing production data & Deliverability forecasts assessed against successive operating cycles \\
\bottomrule
\end{tabular}
\end{table}

\section{Four claims for decision-ready ML}
\label{sec:claims}

\noindent\textbf{Claim 1: uncertainty is the product.} Operators, regulators, and investors need calibrated uncertainty in quantities that determine action: injectivity, pressure footprint, containment, thermal breakthrough, deliverability, and seismicity risk.

Uncertainty arises in geometry, properties, processes and scenarios \cite{mosser2022comprehensive,chen2006data,emerick2013ensemble}. Three recurring types need different treatments. \emph{Epistemic uncertainty} reflects missing knowledge and is handled through priors and updating. \emph{Aleatory uncertainty} is irreducible variability, represented probabilistically and propagated through stochastic simulation. \emph{Interpretational or semantic uncertainty} arises from conflicting interpretations and terminology across disciplines and legacy sources. Probabilistic ML forecasts should report uncertainty in risk-relevant quantities alongside \emph{calibration evidence} and its limits.

Calibration can be evaluated using established diagnostics \cite{gneiting2007strictly,gneiting2014probabilistic}, with engineering interpretations in Table \ref{tab:calibration}. Conformal prediction illustrates why coverage guarantees need to be examined at the level of the intended decision. Standard conformal prediction provides distribution-free, finite-sample marginal coverage under exchangeability of calibration and test data \cite{angelopoulos2023conformal}. This guarantee averages over the randomness in calibration and test samples, so coverage within a particular operating regime can still be poor. Distribution shift can also invalidate the guarantee. Engineering assurance therefore requires decision-specific validation and monitoring of the model's validity domain: the data regimes, geologies and operating conditions for which it has supporting evidence.

\begin{table}[!htbp]
\centering
\small
\caption{Calibration diagnostics for decision-ready uncertainty, and the decision consequence of each.}
\label{tab:calibration}
\begin{tabular}{@{} >{\raggedright\arraybackslash}p{0.25\linewidth} >{\raggedright\arraybackslash}p{0.32\linewidth} >{\raggedright\arraybackslash}p{0.36\linewidth} @{}}
\toprule
\textbf{Diagnostic} & \textbf{What it measures} & \textbf{Decision consequence} \\
\midrule
Reliability diagram / calibration curve & Whether stated probabilities match observed frequencies & Supports traffic-light decisions when calibration is verified at the relevant risk levels and under the intended operating conditions \\
Prediction-interval coverage & Fraction of held-out outcomes falling inside stated intervals & Illustratively, 90\% intervals covering 60\% of held-out pressure data indicate undercoverage; recalibrate uncertainty and reassess pressure-limit exceedance before operational use \\
Continuous ranked probability score (CRPS) & Sharpness and calibration of the full predictive distribution \cite{gneiting2007strictly} & Ranks competing forecast systems for pressure or plume evolution on a proper basis \\
Rank / probability-integral-transform histograms & Empirical rank or PIT distributions relative to uniformity & Assess calibration across operating regimes and evaluate tail-event probabilities directly for seismicity and leakage decisions \\
Expected calibration error (classifiers) & Aggregate miscalibration of event probabilities & Summarises probability calibration; false-alarm and missed-detection rates require threshold-specific evaluation \\
Conformal coverage & Distribution-free finite-sample marginal coverage under exchangeability \cite{angelopoulos2023conformal} & Requires decision-specific validation and validity-domain monitoring; distribution shift can invalidate coverage \\
\bottomrule
\end{tabular}
\end{table}

Subsurface data are \emph{small, biased, heterogeneous, and inconsistent} \cite{bergen2019machine}. Field outcomes are slow and confounded, laboratory and synthetic datasets may not transfer across conditions, and measurement regimes evolve \cite{pawar2015recent,tang2021a}. Combining these data is complicated by incompatible disciplinary schemas and labels, with much of the record unstructured and multi-modal. These limitations make provenance-aware QC essential \cite{wang2025comprehensive}.

\enlargethispage{\baselineskip}
Representing geological uncertainty requires models that encode spatial correlation, stratigraphic constraints, connectivity, and relationships between data types. Structure-aware priors, including implicit fields, graphs, and generative geological models, can reduce data requirements and improve generalisation. They must be paired with validation that detects implausible structures and conditions outside the training distribution~\citep{song2022gansim,di2025latent,shing2017parametrization,varga2019gempy,zhang2019generating,shahri2023a,tang2021a}.

Ensembles~\citep{chen2006data,emerick2013ensemble}, Bayesian surrogates~\citep{lei2021bayesian}, and likelihood-based inversion~\citep{han2025accelerated} should report calibrated predictive intervals that, where possible, separate noise, model error~\citep{wang2025deep}, and scenario uncertainty. Evaluate reported uncertainty using the diagnostics in Table \ref{tab:calibration}, then assess its consequences for recommended actions, monitoring design, or operating envelopes~\cite{mosser2022comprehensive,cao2020a,sun2018discovering,shing2017a}. Where probabilities are poorly constrained, assess whether proposed actions remain acceptable across plausible scenarios. The quantities that drive action should determine model design, observation strategy, and evaluation protocol from the outset.

\noindent\textbf{Claim 2: physics constraints need operational validation.} Physics-constrained models should be evaluated through validity domains, calibration diagnostics, stress tests, and auditable fallbacks for regimes beyond their data or physics assumptions.

``Physics-informed'' approaches include soft regularisation toward governing equations, hard architectural constraints, learned closures embedded in simulators, and wrappers around established models. A model that preserves conservation laws can still give poorly calibrated forecasts under changed conditions. Matching observations can also conceal errors in how the model represents the underlying processes. Methods should preserve the right physical invariants, expose their failure modes, and interact safely with the existing simulation and operational stack.

Hybrid physics-ML approaches can support modelling under data scarcity and shift~\citep{karniadakis2021physics,wang2025comprehensive} by enforcing physical constraints such as conservation, bounds, monotonicity, and stability. Practical safeguards include physics residual checks, constraint violations treated as alarms, and uncertainty-aware fallback to trusted simulators when models extrapolate \cite{rashtbehesht2022physics,he2020physics,fuks2020limitations,almajid2021prediction}.

\emph{Non-intrusive} hybrids use surrogates or inversion wrappers around existing simulators. \emph{Intrusive} hybrids modify solvers through differentiable or adjoint-enabled physics or embedded learned closures~\citep{wang2022surrogate}. Method choice should account for implementation cost and validation under the intended conditions.

\noindent\textbf{Claim 3: benchmarks should evaluate decisions.} A deployment-facing benchmark must ask whether the method changes monitoring placement, narrows uncertainty in pressure-limited operating envelopes, or improves ranking of candidate sites under constraints.

Decision-focused evaluation asks how model errors affect action. Alongside pixel overlap, segmentation evaluation should assess whether segmentation uncertainty materially changes permeability estimates or injectivity forecasts. Alongside mean error, surrogate evaluation should assess whether the surrogate preserves control-strategy rankings or correctly flags pressure-limit exceedance. Section \ref{sec:benchmarks} proposes benchmark designs to evaluate monitoring placement, interval coverage, and operating-envelope quality.

\noindent\textbf{Claim 4: governance is part of the technical stack.} Provenance, failure-mode reporting, calibration tracking, and documentation determine whether ML-enabled recommendations can withstand regulatory and stakeholder scrutiny in public CCS demonstrations, geothermal developments, and H$_2$ storage projects.

ML used in consequential subsurface decarbonisation decisions requires reproducibility, audit trails, stress-testing, and explicit characterisation of failure modes \cite{pawar2015recent}. Adoption also depends on experts in the loop who interrogate outputs, correct interpretations, and update priors with fast feedback. At minimum, adoption requires an auditable record of how data, assumptions, updates, and uncertainty inform decisions \cite{nist2023airmf}. Storage regulation already mandates MMV plans and periodic model updating \cite{eudirective2009,epaclassvi2010}.

Auditability also applies when foundation models, including large language models (LLMs), extract metadata, assumptions, and parameter ranges from reports, scanned logs, and filings to construct priors and scenarios. Every extracted prior, constraint, or parameter range must link to its source passage and receive human verification before entering a decision workflow.

\begin{table}[!htbp]
\centering
\begin{tcolorbox}[
    enhanced,
    fonttitle=\bfseries\large,
    title=Box 1: A decision-readiness checklist,
    colback=white,
    colframe=teal!80!black,
    colbacktitle=teal!15!white,
    coltitle=teal!80!black,
    attach boxed title to top center={yshift=-2mm},
    boxed title style={size=small,colback=white,colframe=teal!80!black},
    boxrule=0.8pt,
    drop fuzzy shadow
]
Proposed checklist for ML outputs that inform consequential operational, regulatory or investment decisions. The eight criteria guide assessment of the supporting workflow and identify gaps in its evidence, validation, oversight and link to action.
\vspace{0.5em}
\begin{enumerate}
\item \textbf{Decision variable named.} The operational or regulatory quantity the output informs (e.g.\ maximum injection rate, sensor placement, trigger threshold) is stated explicitly.
\item \textbf{Uncertainty characterised.} Epistemic, aleatory, and interpretational contributions are distinguished. Probabilistic forecasts report calibration evidence and its limits using relevant diagnostics in Table \ref{tab:calibration}; scenario-based assessments report which actions remain acceptable across plausible scenarios.
\item \textbf{Physical constraints declared.} Conservation laws, bounds, and monotonicity requirements are stated, enforced or checked, and violations are reported as alarms.
\item \textbf{Validity domain stated.} The data regimes, geologies, and operating conditions within which the model is supported are documented, with out-of-distribution detection at inference time.
\item \textbf{Stress tests reported.} Behaviour under distribution shift, sensor dropout, and adversarially chosen but physically plausible scenarios is characterised.
\item \textbf{Provenance and audit trail.} Every output can be traced to data versions, QC decisions, priors, model versions, and human sign-offs (Box 2 gives a worked example).
\item \textbf{Human oversight and fallback defined.} Who can override the model, under what conditions, and what conservative fallback (e.g.\ a trusted simulator or a default envelope) applies when the model abstains.
\item \textbf{Link to action.} The pathway from model output to operational or regulatory action (envelope update, monitoring change, permit evidence) is documented, not implied.
\end{enumerate}
\end{tcolorbox}
\end{table}

\section{Applications where this framing changes practice}

\begin{figure}[!htbp]
    \centering
    \includegraphics[width=\linewidth]{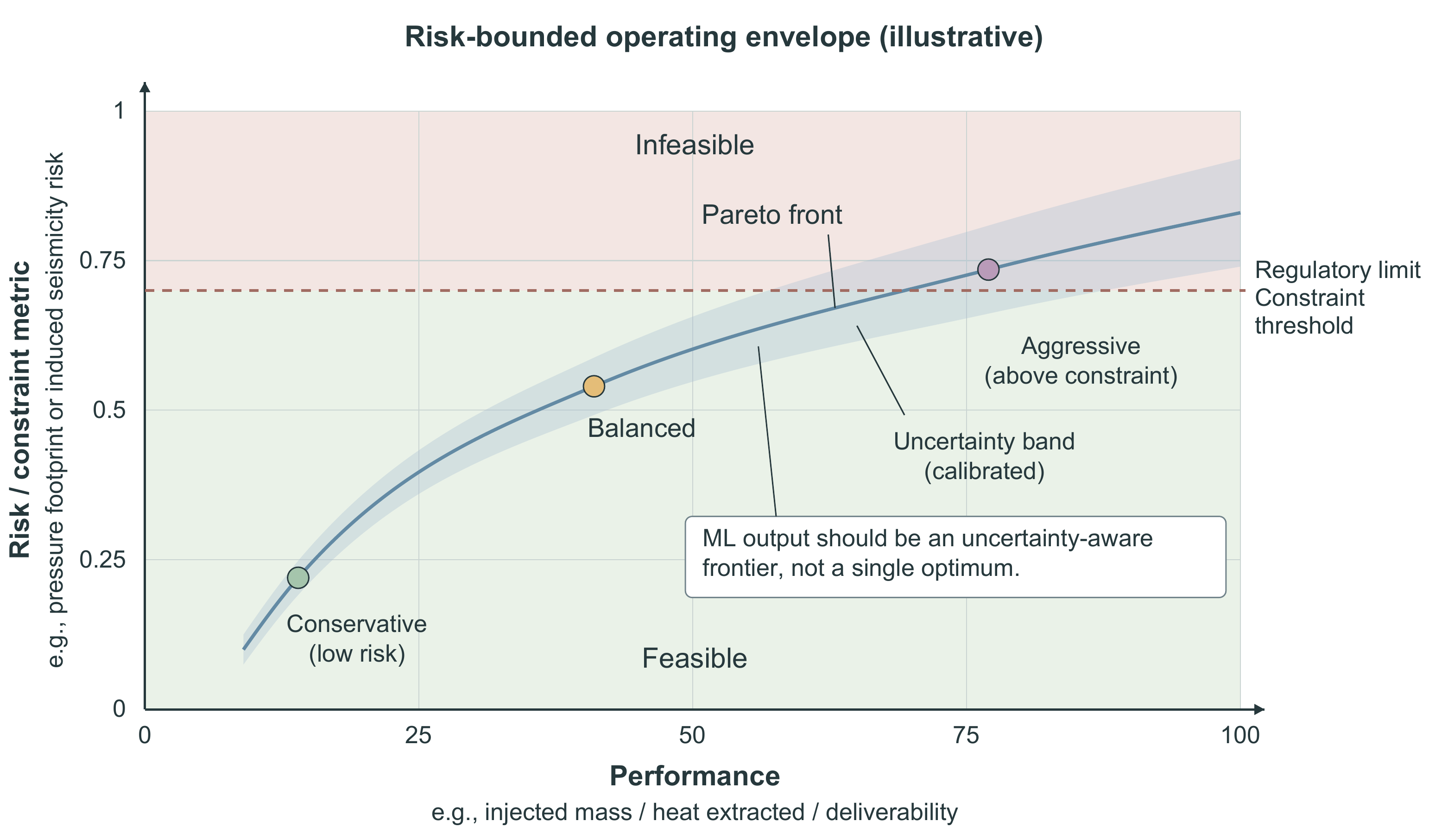}
    \caption{Illustrative decision output for subsurface decarbonisation: an operating envelope / Pareto front showing trade-offs under uncertainty (e.g.\ performance versus pressure footprint and induced seismicity risk). The deliverable is a calibrated, risk-bounded frontier.}
    \label{fig:pareto}
\end{figure}

Figure \ref{fig:pareto} illustrates the target output across these applications: a calibrated operating envelope that makes uncertainty-constrained trade-offs explicit.

\subsection{Pore-to-field integration from imaging and experiments}

Deployment value depends on translating pore-scale mechanisms into field-scale constraints on injectivity decline, permeability evolution, reactive capacity and long-term containment. This requires propagating segmentation and imaging uncertainty through upscaled parameters into decision metrics such as pressure build-up. That propagation remains an outstanding challenge for using pore-scale outputs in appraisal and operations \cite{wu2019predicting,hannah2021upscaling}.

Consistency across pore, core and field scales requires multi-fidelity models, propagation of uncertainty through upscaling, and cross-scale validation \cite{zheng2018adaptive}. Hydro-mechanical feedbacks, including fracture activation, fault slip and stress-dependent permeability, link local operations to far-field outcomes such as induced seismicity and should be represented across scales \cite{meng2026thermo,paluszny2024induced}.

ML can predict effective diffusivity from pore-space images \cite{wu2019predicting}. We propose extending image-based workflows to connect pore structure with multiphase-flow, reactive-transport and mechanical models, and to propagate uncertainty into upscaled parameters. ML opportunities include robust segmentation and feature extraction, learning regime maps (e.g.\ channelling versus uniform dissolution \cite{hannah2023channeling}), accelerating pore-scale simulation \cite{santos2024computationally} for uncertainty studies, and learning closures that preserve conservation and thermodynamic consistency. For H$_2$ storage, pore- and core-scale studies of wettability, hysteresis and microbial activity help constrain models of cyclic field behaviour \cite{heinemann2021hydrogen}.

\subsection{Pressure, trapping, and near-well decisions}
Near-well conformance, relative permeability hysteresis, capillary trapping, and wettability-driven behaviour strongly influence injectivity and storage deliverability, yet are difficult to constrain with limited data \cite{akai2018modeling,ershadnia2020co}. ML can support rate, pressure and cycling decisions that remain acceptable across plausible subsurface states. This support combines fast, uncertainty-aware surrogates for sensitivity screening with inference of effective parameters from sparse laboratory and field observations. Both should operate under explicit physical constraints. Models should transfer across rocks and conditions, and the target deliverable is a calibrated near-well operating envelope (rate or cycling strategy mapped to injectivity and pressure-risk bounds) with clearly defined validity domains \cite{akai2018modeling,ershadnia2020co,mohammadian2022a}.

The same principles apply to control. Reinforcement learning (RL) can support injection-rate, cycling and monitoring decisions by training against physics simulators or hybrid surrogates to search for robust operating rules that respect safety constraints, then validating with conservative stress tests \citep{dixit2023robust}. Outputs should be interpretable operating envelopes, trigger criteria, and contingency actions that remain safe across plausible subsurface realisations.

For porous-media H$_2$ storage with seasonal cycles \cite{heinemann2021hydrogen}, a decision-ready workflow would forecast peak-demand withdrawal rates after $N$ cycles using hysteresis-aware near-well ensembles updated after every cycle. It would report a lower deliverability bound at a probability level chosen for the decision, assess how often held-out outcomes fall below this bound, and flag schedules outside the calibrated validity domain. Forecasts would be scored as each cycle completes.

\subsection{Measurement, monitoring and verification (MMV) and inverse problems}
\label{sec:mmv}
MMV is an inverse problem: infer plume, pressure, and thermal states and leakage risks from incomplete, noisy, multi-modal observations \cite{anikiev2023machine}. At Sleipner, seismic and gravimetric monitoring tracked changes in plume geometry and free CO$_2$ mass \cite{furre2017sleipner}. In injection-driven systems, MMV should also incorporate coupled thermo-hydro-chemical-mechanical processes, using deformation and microseismicity as observations and constraints on safe operating envelopes \cite{pawley2018the,laurenti2022deep}.

Joint inversion and data assimilation translate monitoring observations into estimates of subsurface conditions. Together with uncertainty-aware anomaly detection, these estimates should inform trigger criteria and updated risk bounds for regulators. These estimates should also guide optimal monitoring design, which specifies what to measure, where and when. These methods need explicit treatment of synthetic-to-field gaps and distribution shift. Their success should be judged by earlier detection, fewer false alarms, and reduced uncertainty in risk-relevant quantities, including tail-risk indicators \cite{anikiev2023machine,pawley2018the,laurenti2022deep,tester2021the}.

Ensemble methods, including the ensemble smoother with multiple data assimilation (ES-MDA) and ensemble Kalman filter (EnKF) variants, approximate Bayesian updating in high-dimensional inverse problems~\citep{chen2006data, emerick2013ensemble}. Sequential updates maintain multiple plausible subsurface realisations and propagate uncertainty into risk metrics. ML can provide rapid forecasts and sensitivity analysis through \emph{surrogate forward models}~\citep{wang2022surrogate}. It can also map states to multi-modal data through \emph{learned observation operators} with quantified error, and accelerate posterior updates and what-if analysis through \emph{amortised inversion}. Calibrated uncertainty and stress-testing under shift are essential, especially where the synthetic-to-field gap dominates. Box 2 lists what the audit trail records at each step of one assimilation cycle.

\begin{table}[!htbp]
\centering
\begin{tcolorbox}[
    enhanced,
    fonttitle=\bfseries\large,
    title=Box 2: What an audit trail records in one MMV assimilation cycle,
    colback=white,
    colframe=teal!80!black,
    colbacktitle=teal!15!white,
    coltitle=teal!80!black,
    attach boxed title to top center={yshift=-2mm},
    boxed title style={size=small,colback=white,colframe=teal!80!black},
    boxrule=0.8pt,
    drop fuzzy shadow
]
Illustrative example for a CO$_2$ storage site operating time-lapse seismic and downhole pressure monitoring. Suppose a new pressure series exceeds the ensemble's predicted range. The operator must decide whether to continue under the current envelope, reduce injection, obtain further observations or invoke the site's fallback procedure. Each step records the evidence and reasoning behind that decision.
\vspace{0.5em}
\begin{enumerate}
\item \textbf{Raw observation.} New monitor survey and pressure series archived immutably with acquisition parameters, instrument metadata, and timestamps.
\item \textbf{QC decision.} Processing choices and rejected traces or intervals logged with reasons and the identity of the reviewing analyst; a repeatability metric quantifies survey-to-survey noise.
\item \textbf{Prior state.} The ensemble of subsurface models carried forward from the previous cycle, with version identifiers, its provenance (which data and assumptions produced it), and its predicted observations for the new survey.
\item \textbf{Update.} Assimilation method and settings (e.g.\ ES-MDA iterations, localisation), the surrogate or simulator versions used, and the resulting posterior ensemble, stored alongside the code environment needed to reproduce the update.
\item \textbf{Calibration diagnostic.} Rank histograms and interval coverage of the prior's predictions against the new data (Table \ref{tab:calibration}); a persistently under-dispersed ensemble is flagged before its uncertainty estimates are used downstream.
\item \textbf{Revised risk bounds.} Updated exceedance probabilities for the binding constraints (pressure at the nearest fault, plume arrival at a legacy well), evaluated at current and reduced injection rates over a stated forecast period. Record changes from the previous cycle and which rates meet the stated risk tolerance.
\item \textbf{Human review and action.} Sign-off by the accountable engineer, the rationale linking the risk comparison to the chosen action, any override of the model recommendation with justification, and the resulting action: operating envelope confirmed or revised, monitoring plan amended, or trigger protocol invoked.
\end{enumerate}
\end{tcolorbox}
\end{table}

\enlargethispage{\baselineskip}
MMV workflows should report how new evidence changes operational decisions within safety, regulatory and economic constraints. They should identify which risks have narrowed, which remain unresolved and which measurements would most reduce uncertainty in the next cycle. This requires value-of-information and calibration diagnostics.

Poorly calibrated anomaly detection may generate false alarms, whereas overconfident systems may suppress rare but consequential anomalies. MMV systems need calibrated decision thresholds, quantified false-positive and false-negative trade-offs, and documented human adjudication of ambiguous cases.

A limitation on Claim 3 is clearest in CO$_2$ storage: long-term containment over decades to centuries cannot be evaluated on research timescales. Decision-focused benchmarks can test near-term choices about where to measure, when to intervene, and how to update operating envelopes. Long-run containment claims require prospective forecasts with explicit uncertainty bounds and clear falsification criteria.

\subsection{Basin-scale planning and pore-space competition}

Basin-scale planning involves screening and ranking sites, selecting appraisal programs, and allocating operations and monitoring across interacting projects under shared pressure, seismicity and infrastructure constraints \cite{de2026co2logix,paluszny2024induced}. \emph{Pore-space competition} arises when CCS, geothermal and H$_2$ projects have pressure and stress footprints that interact across licences, operators and timescales \cite{de2026co2logix}. Comparing deployment strategies must account for these interactions and the differing objectives of operators, regulators, network planners and investors. We propose using ML to screen and rank coupled deployment strategies under uncertainty, extending work on pressure-constrained CO$_2$ storage optimisation \cite{liu2025multi}. Such tools should remain coupled to physics and governance constraints \cite{ringrose2016reservoir,pawar2015recent}. Outputs should combine robust rankings and trade-off curves with a conservative, auditable risk envelope.

Basin planning makes subsurface ML an energy-systems coordination task, connecting geological uncertainty with infrastructure timing, network design, and shared environmental constraints. Site value depends on geology and on position relative to CO$_2$ sources, transport corridors, electricity demand, industrial clusters, geothermal heat users, and competing subsurface claims. Regional pressure interference, conflicts with nearby H$_2$ storage, and monitoring liabilities may outweigh a capacity advantage; a more conservative ranking may reduce shared risk and enable faster multi-site build-out.

\section{A near-term agenda for benchmarks, validation, and governance}
\label{sec:agenda}

\subsection{Benchmarks, open datasets, and shared testbeds}
\label{sec:benchmarks}
Existing foundations include the FluidFlower study, which compared blind multi-group simulation forecasts against a physical experiment. For most requested quantities, groups reported non-overlapping P10--P90 uncertainty intervals \cite{flemisch2024fluidflower}. The 11th Society of Petroleum Engineers Comparative Solution Project (SPE11) provides open, precisely specified CO$_2$ storage simulation cases, now broadly taken up by the community \cite{nordbotten2024spe11}. Field-scale infrastructure includes the Bedretto underground laboratory, the Krafla geothermal testbed and the Glasgow Geoenergy Observatories \cite{Ma_BedrettoLab,eichelberger2018krafla,monaghan2022drilling}. The Utah Frontier Observatory for Research in Geothermal Energy (FORGE) provides open geothermal data \cite{moore2019forge}.

Three benchmark designs could add decision-focused evaluation to these foundations:
\begin{itemize}
\item \textbf{Monitoring design.} Given an SPE11-class model ensemble and sensor budget, participants propose monitoring arrays and updating rules. Organisers score these against withheld leakage scenarios using expected detection time, false-alarm rate, and interval coverage of plume and pressure forecasts.
\item \textbf{Pressure envelopes.} Given operational data up to time $t$ from a reference case, participants forecast the maximum sustained injection rate keeping fault-adjacent pressure below a stated limit at 95\% confidence over the next period. Score constraint violations against withheld truth and regret from over-conservatism.
\item \textbf{H$_2$ cycling.} Forecast multi-cycle deliverability with hysteresis using laboratory or pilot data, scored by CRPS and interval coverage per cycle. Outcomes arriving each cycle enable closed-loop scoring on research timescales.
\end{itemize}
Benchmarks should document conditions, uncertainty, and provenance \cite{tang2021a,srinivasan2021a}. Datasets with careful uncertainty annotation should be accompanied by baseline physics models and accessible starter workflows \cite{wang2022surrogate}.

\subsection{Decision-focused validation with limited field evidence}
Evaluation should assess uncertainty reduction in risk-relevant quantities and appropriate changes to recommended actions. Validation should combine cross-validation across modalities, out-of-distribution checks, physics residual and stress tests, and, where feasible, blind prediction challenges \cite{mosser2022comprehensive,karniadakis2021physics,dixit2023robust}. Blind challenges, withheld monitoring campaigns, and prospective forecasts can test model predictions against new observations. FluidFlower illustrates blind comparison against laboratory measurements \cite{flemisch2024fluidflower}. Where synthetic studies stand in for field testing, they should identify which features of the observation operator, geological prior, and operational regime are expected to transfer and which are known simplifications. Remaining uncertainty, reliability boundaries, and failure modes should be visible before operational use.

\subsection{Lessons from adjacent safety-critical fields}
\label{sec:adjacent}
Table \ref{tab:adjacent} summarises what transfers from adjacent safety-critical fields and what does not. Subsurface ML must adapt these calibration and reporting practices to few, slow, and confounded field outcomes. Faster-cycling systems (Table \ref{tab:threesystems}) and closed-loop benchmarks (Section \ref{sec:benchmarks}) are where that adaptation can be tested.

\begin{table}[!htbp]
\centering
\small
\caption{What transfers from adjacent safety-critical fields, and what does not.}
\label{tab:adjacent}
\begin{tabular}{@{} >{\raggedright\arraybackslash}p{0.20\linewidth} >{\raggedright\arraybackslash}p{0.28\linewidth} >{\raggedright\arraybackslash}p{0.22\linewidth} >{\raggedright\arraybackslash}p{0.22\linewidth} @{}}
\toprule
\textbf{Field} & \textbf{Mature practice} & \textbf{Transfers to subsurface} & \textbf{Does not transfer} \\
\midrule
Numerical weather prediction & Continuous ensemble verification; proper scoring rules (CRPS); routine reliability reporting \cite{gneiting2007strictly,gneiting2014probabilistic} & Scoring MMV and deliverability forecasts like weather forecasts; calibration as a reported norm & Verification volume: millions of daily outcomes versus a handful of slow, confounded field outcomes \\
Clinical ML & TRIPOD+AI reporting standard \cite{collins2024tripod}; model cards \cite{mitchell2019model}; device approval pathways & Structured reporting of data, validation, intended use, and failure modes; pre-deployment review & Patient-scale cohorts; randomised evaluation of the decision itself \\
Aviation and nuclear safety & Mandatory incident reporting; shared failure-mode catalogues; safety cases & Standardised failure-mode reporting and safety-case structure for ML workflows & Component-level failure statistics built from large fleets \\
\bottomrule
\end{tabular}
\end{table}

\subsection{Reporting standards and regulatory readiness}
The European Union CO$_2$ storage directive and the United States Environmental Protection Agency Class VI permitting regime require monitoring of storage sites \cite{eudirective2009,epaclassvi2010}. Across CO$_2$ storage, geothermal energy and H$_2$ storage, ML deployment should include traceable evidence, clear human-override responsibilities and explicit validity domains.

The European Union Artificial Intelligence Act sets out risk-management, logging, transparency, and human-oversight requirements for high-risk systems, subject to phased application. Annex III includes AI systems intended to be used as safety components in the management and operation of gas, heating, and electricity supply; classification follows Article 6 \cite{euaiact2024}. The United States National Institute of Standards and Technology Artificial Intelligence Risk Management Framework provides voluntary AI risk-management guidance \cite{nist2023airmf}.

Reporting standards should specify provenance and QC logs, calibration diagnostics, scenario definitions, and model cards describing training data, assumptions, and failure modes \cite{mitchell2019model,pawar2015recent,mosser2022comprehensive}. A subsurface analogue of TRIPOD+AI \cite{collins2024tripod} could define minimum disclosure for decision-facing studies, using Box 1 as a starting point. In regulatory contexts, standards should emphasise traceability from data to decisions, robustness under distribution shift, and conservative extrapolation. Consistent reporting would support comparison and meta-analysis across sites and basins, while enabling reconstruction of the data lineage and uncertainty logic behind recommendations.

\subsection{Government and regulator support}
Many of the highest-value datasets (regional seismic, well logs, monitoring time series, operational data) remain constrained by commercial and regulatory barriers, yet ML progress depends on representative multi-site data and long-duration monitoring records \cite{pawar2015recent,tester2021the}. Governments and regulators can enable controlled access through anonymised releases, data trusts and mandated post-project publication after defined embargoes. They can also fund national and shared testbeds \cite{Ma_BedrettoLab,eichelberger2018krafla,monaghan2022drilling,moore2019forge} and set uncertainty-reporting and audit-trail standards. Public procurement and demonstration funding can require reproducibility, calibration, model governance, and curated post-project datasets in publicly funded CCS, geothermal, and H$_2$ demonstrations, supporting reusable and accountable ML workflows.

Table \ref{tab:mve} summarises actor-specific requirements over an illustrative two-to-three-year horizon.

\begin{table}[!htbp]
\centering
\begin{tcolorbox}[
    enhanced,
    fonttitle=\bfseries\large,
    title=A minimum viable ecosystem: an illustrative near-term horizon,
    colback=white,
    colframe=teal!80!black,
    colbacktitle=teal!15!white,
    coltitle=teal!80!black,
    attach boxed title to top center={yshift=-2mm},
    boxed title style={size=small,colback=white,colframe=teal!80!black},
    boxrule=0.8pt,
    drop fuzzy shadow
]
Actor-specific requirements for moving subsurface decarbonisation ML from isolated case studies into reliable decision workflows.
\vspace{0.5em}

\renewcommand{\arraystretch}{1.4}
\begin{tabular}{@{} >{\raggedright\arraybackslash}p{0.28\linewidth} >{\raggedright\arraybackslash}p{0.68\linewidth} @{}}
\toprule
\textbf{Primary Actor} & \textbf{Ecosystem Requirement} \\
\midrule
\textbf{Academia \& consortia} & \textbf{2--3 open reference testbeds} spanning CCS, geothermal, and/or H$_2$ storage, each with multi-modal data (seismic, wells, core and laboratory, monitoring), clearly defined decision tasks in the style of Section \ref{sec:benchmarks}, and fixed evaluation protocols, building on FluidFlower and SPE11 \cite{flemisch2024fluidflower,nordbotten2024spe11}. Must include at least one testbed with deformation and microseismicity observations. \\
\textbf{Industry \& modellers} & \textbf{A shared uncertainty ``contract''} for reporting geometry, property, and scenario uncertainty (with the calibration diagnostics of Table \ref{tab:calibration}) that is usable in risk metrics and regulator-facing decisions. \\
\textbf{Software vendors \& open source} & \textbf{Interoperable data schemas and open tooling} for end-to-end workflows (ingest $\rightarrow$ QC $\rightarrow$ inversion $\rightarrow$ forecasting $\rightarrow$ decision), including provenance and audit trails (Box 2) by default. \\
\textbf{Publishers \& researchers} & \textbf{Routine decision-focused validation} (stress tests, out-of-distribution checks, physics residual checks, and blind challenges where feasible) embedded into publication and project delivery norms, moving toward a subsurface analogue of TRIPOD+AI \cite{collins2024tripod}. \\
\textbf{Government \& regulators} & \textbf{Policy and procurement requirements} in public CCS, geothermal, and H$_2$ demonstrations: controlled data access, expectations for post-project release, and requirements for reproducibility and uncertainty reporting, aligned with existing storage regulation and emerging artificial intelligence regulation \cite{eudirective2009,epaclassvi2010,euaiact2024}. \\
\textbf{All stakeholders} & \textbf{Standardised failure-mode reporting} and validity domains for common subsurface decarbonisation ML tasks, enabling safe reuse and preventing silent extrapolation beyond calibrated regimes. \\
\bottomrule
\end{tabular}
\end{tcolorbox}
\caption{Minimum viable ecosystem requirements, organised by actor, as an illustrative near-term (two-to-three-year) horizon.}
\label{tab:mve}
\end{table}

\section{Conclusions and next steps}
Machine learning can accelerate subsurface decarbonisation when it is treated as decision infrastructure: shared, persistent, governed systems whose outputs many actors rely on. Deployment requires explicit representation of trade-offs between containment, pressure footprint, induced seismicity, and operational intensity, with uncertainty that can be interrogated at each project stage.

Progress should be judged by whether ML makes appraisal more targeted, operations safer, MMV more informative, and basin-scale deployment more robust. Research consortia should use the proposed pressure-envelope benchmark to compare ML-supported injection decisions with a physics-based baseline, scoring pressure-limit violations and unnecessary restrictions against withheld reference-case results.

\FloatBarrier
\section*{Acknowledgements}
This work is funded by the Engineering and Physical Sciences Research Council's ECO-AI Project grant (reference number EP/Y006143/1).

\section*{Author contributions}
H.P.M. and A.B. conceived the article. H.P.M., A.H.E., L.W., N.W. and A.B. contributed to the scope, structure and writing.

\section*{Competing interests}
The authors declare no competing interests.

\section*{Data availability}
No new data were generated or analysed in this study.

\section*{Code availability}
No custom code was used in this study.

\bibliographystyle{unsrtnat}
\bibliography{references}

\end{document}